\documentclass[11pt,a4paper]{article}
\usepackage{jstyle}
\usepackage{tikz}
\usepackage{amsthm,mathrsfs,stmaryrd,bm}
\usepackage[vcentermath]{youngtab}
\usepackage{simplewick}
\usepackage{framed,pifont}

\newcommand{\tr}{{\rm Tr}}

\def\a{\alpha}
\def\b{\beta}
\def\g{\gamma}

\def\D{\Delta}
\def\e{\epsilon}

\def\l{\lambda}
\def\L{\Lambda}

\def\s{\sigma}

\def\t{\tau}

\author{Euihun JOUNG}
\author{\quad Karapet MKRTCHYAN}
\affiliation{School of Physics and Astronomy, Seoul National University, Seoul 151-747, Korea}

\emailAdd{euihun.joung@snu.ac.kr}
\emailAdd{karapet@snu.ac.kr}

\title{\centering
\LARGE{Partially-massless higher-spin algebras\\
and their finite-dimensional truncations}}

\abstract{The global symmetry algebras of partially-massless (PM) higher-spin 
(HS) fields in (A)dS$_{d+1}$ are studied.
The algebras  involving PM generators up to depth $2\,(\ell-1)$
are defined as the maximal symmetries of free conformal scalar field with $2\,\ell$ order wave equation in $d$ dimensions. 
We review the construction of these algebras by quotienting certain ideals in the universal enveloping algebra of $(A)dS_{d+1}$ isometries.
We discuss another description in terms of Howe duality and derive the formula for computing trace in these algebras.
This enables us to explicitly calculate the bilinear form for this one-parameter family of algebras.
In particular, the bilinear form shows the appearance of additional ideal for any non-negative integer values of $\ell-d/2$\,, which coincides with the annihilator of the one-row $\ell$-box Young diagram representation of $\mathfrak{so}_{d+2}$\,.
Hence, the corresponding finite-dimensional coset algebra spanned by massless and PM generators is equivalent to the symmetries of this representation.}

\begin{document}

\maketitle

\section{Introduction}
\label{sec: intro}

Higher spin (HS) symmetries play a central role in the consistency of HS theories 
\cite{Fradkin:1986ka,Fradkin:1986qy,Vasiliev:1990en,Vasiliev:2003ev,Bekaert:2005vh},
and serve as important guideline in the understanding of different AdS/CFT models.
In the recent work \cite{Joung:2014qya} the authors of the current article have studied several aspects --- such as minimal coadjoint orbit, minimal representation and Joseph ideal --- related to symmetries
of massless HS fields of symmetric index type, and found a convenient formula for computing trace in HS algebra.
In this article, we continue our previous work and clarify analogous aspects 
of a more general type of HS symmetries ---
global symmetries of a theory involving partially-massless (PM) HS fields
\cite{Deser:2001us}. 

PM fields are special spectra which exist (or stay irreducible) only in 
the backgrounds with non-vanishing constant curvature. More precisely, they are unitary 
irreducible in dS background, whereas in AdS they are irreducible but not unitary.
For a given spin $s$, there are $s$ different PM fields labelled by their \emph{depth} $t=0,1,\ldots,s-1$\,,
where $t=0$ corresponds to the massless field.
They are described by gauge potentials that transform with gauge parameters 
of lower rank compared to massless fields of the same spin:
\be
	\delta\,\varphi^{\sst (t)}{}_{\mu_{1}\cdots\mu_{s}}
	=\nabla_{(\mu_{1}}\cdots\nabla_{\mu_{t+1}\phantom{|}}\varepsilon^{\sst (t)}{}_{\mu_{t+2}\cdots \mu_{s})}
	+\mathcal O(\Lambda)\,,
\ee
where $\varphi^{\sst (t)}$ is a PM field of depth $t$, $\nabla$ denotes (A)dS covariant derivative
and $\L$ is the cosmological constant. 
The propagating degrees of freedom of the
spin $s$ PM field with depth $t>0$ are more than those of the massless spin $s$ but less than those of the massive one. 
In the flat space limit, PM field of depth $t$ is decomposed into a collection of massless fields of spin \mt{s, s-1,\ldots, s-t}\,\cite{Zinoviev:2001dt}. 
The Killing tensor of a PM field of spin $s$ and depth $t$ are given by $O(d+2)$ Young diagram with two rows of respective length $s-1$ and $s-1-t$ \cite{Skvortsov:2006at}.
Therefore, algebras involving such generators can be interpreted
as global symmetry of a theory involving PM fields. 

In order to construct a physically consistent algebra
based on these generators, we need to require
following two conditions:
\begin{itemize}
\item The isometry algebra $\mathfrak{so}_{d+2}$\footnote{In this article, we do not consider any issue related to the reality structure.} is a subalgebra.
This implies that the corresponding HS theory contains gravity sector.

\item All the generators of different spins transform covariantly under isometries.
This is to say that in the corresponding theory, HS fields couple to gravity in a diffeomorphism invariant manner, which is tantamount to the requirement of the equivalence principle to hold.
\end{itemize}
A priori, finding out an algebra satisfying the above conditions is highly non-trivial task
in a bottom-up approach,  as we have to solve for the Lie brackets among 
different PM generators (labelled by infinitely many spins and depths) satisfying Jacobi identity\footnote{These kind of construction was discussed recently in \cite{Boulanger:2013zza}.}. Fortunately, 
a series of algebras involving PM HS generators and 
satisfying the above conditions have been already found in the literature. 
The first discussions are from mathematics literature, 
the simplest case \cite{Eastwood:2006}
and the generalizations \cite{Gover:2009, Michel:2011},
where certain PM HS algebras are defined as the maximal symmetries of higher order Laplace operators
\be
	\Box^{\ell}\,\phi=0\,,
	\label{ho single}
\ee
generalizing the definition of the massless HS algebra ($\ell=1$)\,. 
In physics, the unfolded formulation of the corresponding PM HS spectra
has been proposed in \cite{Bekaert:2005vh}
and  explored in \cite{Alkalaev:2007bq,Alkalaev:2011zv}.
In \cite{Bekaert:2012vt,Bekaert:2013zya,Basile:2014wua}, the Flato-Fronsdal theorem was generalized to higher-order singletons and arguments were provided for a conjecture of AdS/CFT duality, relating a Vasiliev-type theory with PM symmetric tensor fields to the $O(N)$ model at a multicritical isotropic Lifshitz point.

In this paper, we revisit the PM algebras
including the aforementioned ones 
from various points of view. 
We first define the PM algebras as particular cosets of
universal enveloping algebra (UEA) of the isometry algebra. 
This requires to identify proper ideals of the UEA
and such ideals depend on a parameter $\lambda$\,.
For generic value of $\l$\,, the algebra
is spanned by PM generators of any even depths.
When $\lambda$ takes an integer value $\ell$\,, 
it truncates into the algebra associated with \eqref{ho single}
allowing only the generators of depth smaller than $2\,\ell$\,.
We also show how the same algebras can be constructed
making use of oscillators and Howe duality. 
The oscillator description allows us to perform more concrete calculations 
and we derive the explicit form of trace  and bilinear form for these PM algebras.
The explicit expression of bilinear form reveals 
a new structure of PM algebras.
For (half-)integer values of $\l$ greater than $d/2$\,, with (odd)even $d$,
the algebras truncate into
\emph{finite dimensional} ones.

The paper is organized as follows. In Section \ref{2}, we provide generalities on the global symmetries of free PM HS fields and the vector space of PM HS algebra. In Section \ref{3}, we define PM HS algebras as cosets of 
the UEA of (A)dS isometry algebra.
In Section \ref{4}, we present Howe duality and oscillator realization of PM HS algebras.
In Section \ref{5}, the trace for the PM HS algebras are defined
and the bilinear forms are calculated.
In Section \ref{6}, we show
the finite-dimensional truncation of algebras for special values of $\l$. 
In Section \ref{7}, the relations of the PM HS algebras to 
conformal higher spin theory are discussed.
Appendix \ref{A} contains computational details.

\section{Partially-Massless Fields and Their Global Symmetries}
\label{2}

The global-symmetry generators, or Killing tensors, corresponding to PM fields can be
identified by analyzing the relevant Killing equations. 
They were first identified in the frame-like formulation in \cite{Skvortsov:2006at}.
Below, we shall rederive the PM Killing tensors in the metric-like ambient formulation.
As it has been shown in \cite{Alkalaev:2011zv,Joung:2012rv},
the Fierz system of the PM field of spin $s$ and depth $t$
has the following gauge equivalence relation,
\be
	\delta_E\,\Phi^{\sst (t)}(X,U)=(U\cdot \partial_{X})^{t+1}\,E^{\sst (t)}(X,U)\,,
\ee
where PM gauge parameters $E^{\sst (t)}$ are traceless tensors:
\be
	\partial_{U}^{2}\,E(X,U)=0\,.
	\label{traceless}
\ee
As usual, being ambient-space fields, $\Phi^{\sst (t)}$ and $E^{\sst (t)}$
should satisfy the tangentiality and homogeneity conditions:
\ba
	&X\cdot\partial_{U}\,\Phi^{\sst (t)}(X,U)=0\,,
	\qquad
	&\left(X\cdot\partial_{X}-U\cdot\partial_{U}+2+t\right)\,\Phi^{\sst (t)}(X,U)=0\,,\nn
	&X\cdot\partial_{U}\,E^{\sst (t)}(X,U)=0\,,
	\qquad
	&\left(X\cdot\partial_{X}-U\cdot\partial_{U}-t\right)\,E^{\sst (t)}(X,U)=0\,,
\ea
in order to be reduced to (A)dS Fierz system.
The global symmetries are given by parameters, that satisfy the Killing equation,
\mt{\delta_E\, \Phi^{\sst (t)}=0}\,, which combined with the tangentiality and homogeneity conditions
defines the following system,\footnote{One 
can equally describe PM fields in the Stueckelberg formulation where the gauge 
transformation has the standard form,
\be
	\delta_E\,\Phi^{\sst (t)}(X,U)=U\cdot \partial_{X}\,E^{\sst (t)}(X,U)\,,
\ee
whereas the tangentiality condition is modified to
\be
	(X\cdot\partial_{U})^{t+1}\,E^{\sst (t)}(X,U)=0\,.
\ee
Therefore, in the Stueckelberg formulation, we get conditions
equivalent to \eqref{PM Killing} but $X$ and $U$ exchanged.}
\be
	X\cdot\partial_{U}\,E^{\sst (t)}=0\,,
	\quad
	(U\cdot\partial_{X})^{t+1}\,E^{\sst (t)}=0\,,
	\quad
	(X\cdot\partial_{X}-U\cdot\partial_{U}-t)\,E^{\sst (t)}=0\,.
	\label{PM Killing}
\ee
It is obvious to see from here that the PM Killing tensors of depth $t$ correspond to the
$(t+1)$-dimensional representation of $\mathfrak{sp}_{2}$\,:
\ba
	&[h,e]=2\,e\,,\qquad [h,f]=-2\,f\,,\qquad  [e,f]=h\,,&\nn
	&e=U\cdot\partial_X\,,
	\qquad f=X\cdot\partial_U\,,\qquad
	h=X\cdot\partial_X-U\cdot\partial_U\,.&
\ea
We shall also see the relevance of this $\mathfrak{sp}_{2}$ representation later in
the other construction.
The conditions \eqref{PM Killing} and \eqref{traceless} are solved by
\be
	E^{\sst (t)}(X,U)=\sum^{\infty}_{r=0}
	\frac1{r!\,(r-t)!}\,X^{\sst M_{1}}\cdots X^{\sst M_{r}}\,
	U^{\sst N_{1}}\cdots U^{\sst N_{r-t}}\,
	M^{\sst (r,t)}_{\sst M_{1}\cdots M_{r},N_{1}\cdots N_{r-t}}\,,
\ee
with the tensor $M^{\sst (r,t)}$ taking values in the $\{r,r-t\}$ Young diagram of $O(d+2)$\,:
\be
	M_{\sst M_{1}\cdots M_{r},N_{1}\cdots N_{r-t}} \sim\,
	\parbox{145pt}{\begin{tikzpicture}
	\draw (0,0) rectangle (5,0.4);
	\draw (0,0) -- (0,-0.4) -- (3,-0.4) -- (3,0);
	\draw [dotted] (5,0) -- (5,-0.4) -- (3,-0.4);
	\draw [very thin, gray] (0.4,0.4) -- (0.4,-0.4);
	\draw [very thin, gray] (4.6,0.4) -- (4.6,0);
	\draw [very thin, gray] (2.6,-0.4) -- (2.6,0);
	\node at (2.5,0.2){${\st r}$};
	\node at (1.5,-0.2){$\st r-t$};
	\node at (4,-0.2){$\st t$};
	\end{tikzpicture}}\,.
	\label{PM gen}
\ee
Therefore, any theory involving a PM field of spin $s$ and depth $t$ 
should admit global symmetry
containing generator $M^{\sst (s-1,t)}$\,.

\section{Coset Construction from Universal Enveloping Algebra}
\label{3}

Similarly to the massless HS algebra, 
the algebras involving PM HS generators can be approached from the universal enveloping algebra (UEA)
of $\mathfrak{so}_{d+2}$\,.
One of the advantages of UEA construction is that the two physical consistency conditions mentioned in Introduction are automatically satisfied.
Let us remind that the massless HS algebra can be obtained as the coset,
\be
	hs(\mathfrak{so}_{d+2})=\mathcal U\ \Big/
	\left({\tiny \yng(2)}\oplus\,{\tiny{\yng(1,1,1,1)}}\ \right),
\ee
where $\mathcal U$  denotes the UEA of $\mathfrak{so}_{d+2}$,
while (---) denotes the Joseph ideal, generated by the following elements in
$\mathfrak{so}_{d+2}\odot\mathfrak{so}_{d+2}$\,:
\be
	J_{ab}:= M_{(a}{}^{c}{}\odot M_{b)c}-\frac{\eta_{ab}}{d+2}\,
	M^{cd}\odot M_{cd}\sim {\tiny\yng(2)}\,,
	\qquad
	J_{abcd} := M_{[ab}\odot M_{cd]}
	\sim {\tiny\yng(1,1,1,1)}\ .
\ee
This procedure determines the eigenvalues of all Casimir operators of $\mathfrak{so}_{d+2}$.
In particular, the quadratic Casimir is given by
\be
	C_{2}:=\frac12\,M_{ab}\odot M^{ba}=-\frac{(d+2)(d-2)}4\,.
\ee
Let us now consider deformations of the above construction
where we take a quotient with the ideal
generated by only one of $J_{ab}$
 and $J_{abcd}$\,.
 In such cases, the quadratic Casimir $C_{2}$ remains arbitrary while the other Casimirs
are fixed as functions of $C_{2}$ as shown in \cite{Iazeolla:2008ix,Boulanger:2011se}.
Hence, one can take a further quotient with $C_{2}-\nu$\,.
In this way, we have two ideals labelled by a continuous parameter $\nu$ as
\be
	\mathcal I(\nu)=
	\left( \, {\tiny{\yng(1,1,1,1)}}\ \oplus (C_{2}-\nu) \right),
	\qquad
	\mathcal J(\nu)=
	\Big( \,{\tiny{\yng(2)}} \oplus (C_{2}-\nu) \Big)\,.
	\label{ideals}
\ee
Since these ideals contain fewer elements
than the Joseph ideal,
the corresponding coset algebras contain more generators than the original one. These one-parameter
families of algebras have already been investigated in
\cite{Michel:2011} and \cite{Boulanger:2011se}, respectively. 
In the case of $\mathcal I(\nu)$\,, we get the global symmetry algebra of
a theory involving PM HS fields, 
whereas for $\mathcal J(\nu)$\,, we get the symmetry of 
a theory involving mixed symmetry HS fields.
Leaving aside the discussion about the mixed symmetry case,
in the current article we shall consider only the ideal $\mathcal I(\nu)$ and corresponding PM HS algebras.

Let us resume the discussion with the coset algebra,
\be\label{Anu}
 	\mathcal A_{\lambda}=\mathcal U/\,
	\mathcal I(\nu_{\lambda})\,,
	\qquad
	\nu_{\lambda}=-\frac{(d-2\,\lambda)(d+2\,\lambda)}{4}\,,
\ee
where we have parametrized $\nu$ in terms of $\lambda$
for later convenience.
The ideal generated by $J_{abcd}$ contains all $GL(d+2)$-tensors
in $\mathcal U$ having more than two rows, hence
the algebra $\mathcal A_{\lambda}$ is spanned by the generators
which have the symmetry of two-row $GL(d+2)$ Young diagrams.
When decomposed into the traceless $O(d+2)$ tensors, it is
given by
\be
	\mathcal A_{\lambda}\simeq \bigoplus_{p=0}^{\infty}\,
	K_{p}\,,
	\label{PMdecomposition}
\ee
where $K_{p}$ are the vector space of
the Killing tensors associated with the PM field
of any spin and depth \mt{t=2p}\,:
\be
	K_{p}=\bigoplus_{r=0}^{\infty}\
	\parbox{145pt}{\begin{tikzpicture}
	\draw (0,0) rectangle (5,0.4);
	\draw (0,0) -- (0,-0.4) -- (3,-0.4) -- (3,0);
	\draw [dotted] (5,0) -- (5,-0.4) -- (3,-0.4);
	\draw [very thin, gray] (0.4,0.4) -- (0.4,-0.4);
	\draw [very thin, gray] (4.6,0.4) -- (4.6,0);
	\draw [very thin, gray] (2.6,-0.4) -- (2.6,0);
	\node at (2.5,0.2){${\st r+2p}$};
	\node at (1.5,-0.2){$\st r$};
	\node at (4,-0.2){$\st 2p$};
	\end{tikzpicture}}\,.
\ee
Hence,  the generators of $\mathcal A_{\lambda}$
are not only the massless Killing tensors $K_{0}$
but also the PM ones $K_{p\ge1}$ corresponding to the depth $2p$ 
\cite{Alkalaev:2007bq} (see also \cite{Alkalaev:2014nsa} for related discussions).

As has been shown in \cite{Michel:2011}\,,
when $\lambda$ takes integer values $\ell
=1,2,\ldots,$
the algebra $\mathcal A_{\ell}$ develops the ideals $\mathfrak{q}_{\ell}$
which consist of PM Killing tensors of even depth not smaller than $2\,\ell$\,:
\be
	\mathfrak{q}_{\ell} \simeq
	\bigoplus_{p=\ell}^{\infty} K_{p}\,.
\ee
As a result, the algebra is decomposed into
\be
	\mathcal A_{\ell}=\mathfrak{p}_{\ell}
	\oplus
	\mathfrak{q}_{\ell}\,,
\label{Aell}
\ee
where $\mathfrak{p}_{\ell}$ is the coset algebra:
\be
	\mathfrak{p}_{\ell}=\mathcal A_{\ell}/\mathfrak{q}_{\ell}\simeq
	\bigoplus_{p=0}^{\ell-1} K_{p}\,,
	\label{vect p}
\ee
containing generators of even depth smaller than $2\,\ell$\,.
Therefore, we obtain two algebras $\mathfrak{q}_{\ell}$ and $\mathfrak{p}_{\ell}$ for $\lambda=\ell$\,: 
the former algebra contains generators of even depths not smaller than $2\,\ell$\,\footnote{If realized field theoretically, $\mathfrak{q}_{\ell}$ should correspond to  global symmetry of an exotic interacting theory involving
PM fields of even depths not smaller than $2\,\ell$\,.},
whereas the latter one is associated with the theory
involving PM fields of even depths smaller than $2\,\ell$\,.
The usual massless HS algebra corresponds to $\mathfrak{p}_1$\,.

Let us focus now on the coset algebra $\mathfrak{p}_{\ell}$
which contains the usual HS symmetry in the $\ell=1$ case.
In fact, it can be obtained directly from $\mathcal U$ as the coset,
\be
	\mathfrak{p}_{\ell}=\mathcal U/\,\mathcal I_{\ell}\,,
	\label{coset p}
\ee
where the ideal $\mathcal I_{\ell}$ is
given by \cite{Gover:2009}
\be
 	\mathcal I_{\ell}=\left(
	\parbox{74pt}{\begin{tikzpicture}
	\draw (0,0) rectangle (2.5,0.26);
	\draw [very thin, gray] (0.26,0) -- (0.26,0.26);
	\draw [very thin, gray] (2.24,0) -- (2.24,0.26);
	\node at (1.25,0.14) {$\st 2\,\ell$};
	\end{tikzpicture}}
	\oplus\,{\tiny{\yng(1,1,1,1)}}\ \right).
	\label{I ell}
\ee
The vector space \eqref{vect p} of the $\mathfrak{p}_{\ell}$ can be 
straightforwardly identified using this ideal,
or equivalently from the corresponding coadjoint orbit defined by
\be
	W^{(a_{1}}{}_{b_{1}}\,W^{c_{1}}{}_{d_{1}}\cdots\,
	W^{a_{\ell}}{}_{b_{\ell}}\,W^{c_{\ell})}{}_{d_{\ell}}\,
	\eta^{b_{1}d_{1}}\,\cdots \eta^{b_{\ell}d_{\ell}}=0\,,
	\qquad
	W^{a[b}\,W^{cd]}=0\,.
	\label{pm orbit}
\ee	
Any polynomials of $W^{ab}$ subject to the above conditions
are dual to the generators of $\mathfrak p_{\ell}$
so give rise to the vector space \eqref{vect p}\,.
The values of the quadratic Casimir \eqref{Anu}
can be also calculated from the ideal $\mathcal I_{\ell}$ by
inquiring the consistency of the ideal.

The algebra $\mathfrak{p}_{\ell}$ corresponds in fact to the symmetry algebra of the $\ell$-th power of
Laplacian operator, or in other words the symmetry of the action,
\be\label{boxell}
	S[\phi]=\int d^{d}x\,\phi\,\Box^{\ell}\,\phi\,.
\ee
This symmetry has been studied for $\ell=1,2$ in
\cite{Eastwood:2002su, Eastwood:2006}
and for arbitrary positive integers $\ell$ in  \cite{Gover:2009}.
The aspect of holographic correspondence has been
also explored recently in \cite{Bekaert:2013zya}.

For generic values of $\lambda$\,, the algebra $\mathcal{A}_\l$ 
is generated by the Killing tensors of all even depth PM HS fields. 
Hence, if there exists a theory based on this field content, then $\mathcal{A}_\l$ may well
correspond to its global symmetry algebra. 
However, the viability of such a theory is not obvious:
since the value of the quadratic Casimir used for the quotienting does not correspond any more to
that of a free scalar field on the boundary, it is not clear what might be the CFT dual.
The ideal $\mathfrak{q}_\ell$ appearing for the integer value of $\l=\ell$ corresponds to an (higher-order) on-shell condition 
for the boundary scalar,
but $\mathcal{A}_{\l}$ does not admit such an ideal, hence the boundary scalar seems to be an off-shell one with a fixed conformal weight. It will be interesting to examine the relevant Flato-Fronsdal theorem
to see whether this  CFT might be dual to the putative PM HS theory, but it is beyond the scope of the current paper.\footnote{We are indebted to the anonymous referee for pointing out this subtle point.}

\section{Howe Duality and Oscillator Realization}
\label{4}
The coset construction from UEA provides a concise definition of the PM symmetries,
but for explicit calculations we may consider another description of PM algebra. 
Similarly to massless case, PM algebra $\mathcal{A}_{\lambda}$ can be 
defined using a reductive dual pair correspondence (introduced in HS context in \cite{Vasiliev:2003ev}), aka Howe duality, as
\be
	\big(\,Sp(2)\,,\,O(d+2)\,\big)\subset Sp(2(d+2))\,.
\ee
The starting point is again the metaplectic representation of $\mathfrak{sp}_{2(d+2)}$
which can be realized by the star product algebra with the product,\footnote{As
 in \cite{Joung:2014qya}, we denote $\mathfrak{sp}_2$ indices by greek letters and $\mathfrak{so}_{d+2}$ indices by roman letters. The $\mathfrak{sp}_2$ indices are rised and lowered using antisymmetric invariant tensors $\epsilon_{\a\b}$ and $\epsilon^{\a\b}$, while $\mathfrak{so}_{d+2}$ indices are rised and lowered by symmetric metric tensor of $\mathfrak{so}_{d+2}$ (which can be taken as Kroneker delta, since we work with the complex algebra, not a particular real form) and its inverse.}
\be
	(f\star g)(y)
	=\exp\Big(\frac12\,\e_{\a\b}\,\partial_{y_{\a}}\!\cdot \partial_{z_{\b}}\Big)\,
	f(y)\,g(z)\,\Big|_{z=y}\,.
\ee
The mutual stabilizers $\mathfrak{sp}_2$ and $\mathfrak{so}_{d+2}$ correspond to
\be
	K_{\a\b}=y_{\a}\!\cdot y_{\b}\,,
	\qquad 
	M_{ab}=y_{\a a}\,y^{\a}{}_{b}\,,
	\label{M rep}
\ee
and the off-shell HS algebra  $\widetilde{hs}(\mathfrak{so}_{d+2})$ 
is defined as the centralizer of $\mathfrak{sp}_2$\,:
any element $f(y)$ in $\widetilde{hs}(\mathfrak{so}_{d+2})$ satisfies
\be
	[\,f(y)\ \overset{\star},\  K_{\a\b}\,]=0\,.
\ee
By solving this condition, one can show that $\widetilde{hs}(\mathfrak{so}_{d+2})$ 
is spanned by polynomials of $M_{ab}$\,.

The usual (on-shell) massless HS algebra is 
the coset of $\widetilde{hs}(\mathfrak{so}_{d+2})$ 
by the $\mathfrak{sp}_2$-triviality relation, $K_{\a\b}\sim0$\,.
This relation actually defines the trivial representation of   $\mathfrak{sp}_2$\,,
which is dual to the singleton representation of $\mathfrak{so}_{d+2}$\,.
In the following, we shall consider generalization
of the last step and define the PM algebras $\mathcal A_{\l}$
and $\mathfrak p_{\ell}$\,.

\paragraph{PM algebra $\bm{\mathcal A_{\l}}$}
In order to obtain $\mathcal A_{\lambda}$ from $\widetilde{hs}(\mathfrak{so}_{d+2})$\,,
we can quotient the latter with 
the $\mathfrak{sp}_{2}$ Casimir relation,
\be
	C_{2}(\mathfrak{sp}_2)=
	\frac12\,K_{\a\b}\star K^{\a\b}
	\sim (1-\lambda)(1+\lambda)\,,
	\label{soCasimir}
\ee
which also fixes the quadratic Casimir of $\mathfrak{so}_{d+2}$ 
in terms of $\lambda$ to \eqref{Anu}
since in $\widetilde{hs}(\mathfrak{so}_{d+2})$ the two Casimirs are related by
\be
	K_{\a\b}\star K^{\a\b}
	+ M_{ab}\star M^{ba}=-\frac{(d+2)(d-2)}{2}\,.
\ee
In the case of $\lambda=\ell$\,,
the relation \eqref{soCasimir}  
gives an indecomposable representation
consisting of the $(\ell+1)$-dimensional representation
and the infinite-dimensional representation with lowest (or heighest) weight
$\ell+1$ (or $-(\ell+1)$)\,.
Hence, the algebra $\mathcal A_{\ell}$
becomes the semi-direct-sum of  the coset algebra $\mathfrak{p}_{\ell}$
and the ideal algebra $\mathfrak q_{\ell}$ as in \eqref{Aell}.

\paragraph{PM algebra $\bm{\mathfrak{p}_{\ell}}$}

In order to obtain directly $\mathfrak{p}_{\ell}$ from $\widetilde{hs}(\mathfrak{so}_{d+2})$\,,
we can quotient the latter with the following equivalence relations 
\be\label{ell rep}
	K_{++}\sim 0\,,
	\quad
	K_{--}^{\, \ell} \sim 0\,,
	\quad
	\left(K_{+-}+\ell-1\right) \sim 0\,,
\ee	
which define the $(\ell+1)$-dimensional representation
of $\mathfrak{sp}_2$\,.
This relations have asymmetric form with respect to $\mathfrak{sp}_2$ indices, but
one can show in fact that any other relations defining the $(\ell+1)$-dimensional representation
such as,
\be
	K_{++}^{\,1+k}\sim 0\,,
	\quad
	K_{--}^{\,\ell-k} \sim 0\,,
	\quad
	\left(K_{+-}+\ell-k-1\right) \sim 0\,,
	\qquad[\,k=0,1,\ldots,\ell\,]\,,
	\label{PM equiv}
\ee
give the same algebra $\mathfrak{p}_\ell$.
Instead of \eqref{PM equiv},
one may also use the equivalence relation\footnote{The same $\mathfrak{sp}_{2}$ tensors have been used in decomposing generators of off-shell HS algebra into PM ones in \cite{Alkalaev:2007bq}.} 
\be
K_{(\a_1\a_2}\star K_{\phantom{(}\a_3\a_4}\star
\cdots\star K_{\a_{2\ell-1}\a_{2\ell})}\sim 0\,,\label{PM covariant condition}
\ee
which generalizes the relation $K_{\a\b}\sim 0$ of the massless HS algebra
to PM cases in a manifestly $\mathfrak{sp}_2$ covariant manner.
However, this relation is weaker than \eqref{PM equiv}
since the relations $K_{(\a_{1}\a_{2}}\star\cdots\star K_{\a_{2n-1}\a_{2n})}\sim0$
with $n=0,1,\ldots,\ell-1$ also imply \eqref{PM covariant condition}.
Therefore, only complemented by \eqref{soCasimir}, 
the relation
\eqref{PM covariant condition}
becomes equivalent to \eqref{PM equiv}
so defines the PM algebra $\mathfrak{p}_{\ell}$\,. The condition \eqref{soCasimir} implies multiplicity one for each generator of given spin and depth.
In the following we shall use the relations \eqref{soCasimir}
and \eqref{PM covariant condition}
to derive the trace of PM algebras.
Instead, in Appendix \ref{A}, we make use of relations \eqref{ell rep} for alternative derivation of the trace.

\subsection*{Decomposition of traceful generators into PM ones}

For concreteness, let us look at
the generators of $\mathcal A_{\l}$ 
(and $\mathfrak{p}_{\ell}$) more closely
and show that they indeed correspond to PM Killing tensors given in \eqref{PM gen}.
We begin with $\widetilde{hs}(\mathfrak{so}_{d+2})$ spanned by
\be
	\tilde M^{\sst (r)}_{a_{1}\cdots a_{r},b_{1}\cdots b_{r}}
	=y_{a_{1}\a_{1}}\,y_{b_{1}}{}^{\a_{1}}\,\cdots\,y_{a_{r}\a_{r}}\,y_{b_{r}}{}^{\a_{r}}\,,
	\label{tf tensor}
\ee
which are traceful tensors corresponding to $GL(d+2)$ $\{r,r\}$ Young diagrams.
The traceful tensor $\tilde M^{\sst (r)}$ \eqref{tf tensor} can be decomposed into traceless tensors $M^{\sst (r,2t)}$
with $t=0,\ldots, [r/2]$\,,
\be
	M^{\sst (r,2t)}_{a_{1}\cdots a_{r},b_{2t+1}\cdots b_{r}}
	=\tilde M^{\sst (r)}_{a_{1}\cdots a_{r},b_{1}\cdots b_{r}}\,
	\eta^{b_{1}b_{2}}\cdots \eta^{b_{2t-1}b_{2t}}
	-({\rm traces})\,,
\ee
which can be interpreted as PM generators of spin $r+1$ and depth $2t$\,.
These tensors can be conveniently handled by contracting them with
\be
	\tilde W^{ab}=\tilde w^{a}_{\a}\,\tilde w^{\a b}\,.
	\label{W para}
\ee
Then, the traceless generators of $\widetilde{hs}(\mathfrak{so}_{d+2})$ given by
\be
	M^{\sst (r,2t)}(\tilde W)
	= M^{\sst (r,2t)}_{a_{1}\cdots a_{r},b_{2t+1}\cdots b_{r}}\,
	\eta_{b_{1}b_{2}}\cdots \eta_{b_{2t-1}b_{2t}}\,
	\tilde W^{a_{1}b_{1}}\cdots \tilde W^{a_{r}b_{r}}\,,
	\label{M W}
\ee
can be expressed solely through  traces of $M_{ab}$'s and $\tilde W^{ab}$'s,
\be
	\la M^{p}\,\tilde W^{q}\,M^{r}\,\cdots\,\tilde W^{s}\ra\,,
	\label{traces}
\ee
as they do not have any free index.
Using the following identity based on \eqref{M rep} and \eqref{W para},
\be 
	A\,B\,A=\frac12\,A\,\la A\,B\ra\,,
	\qquad [\,A,B=M\ {\rm or}\ \tilde W\,],
	\label{M W id}
\ee
one can show that any trace of type \eqref{traces} will
be broken down to a function of
\be
	\la M^{2}\ra \la \tilde W^{2}\ra\,,
	\qquad
	\la M\,\tilde W\ra\,,\qquad \la (M^{2}\,\tilde W^{2})^{p}\ra\,.
\ee
Here again, one can show that the last object with  $p\ge2$
is not independent and can be expressed in terms of the rest.
Hence, $M^{\sst (r,2t)}(\tilde W)$ in \eqref{M W} are polynomials of $\la M^{2}\ra \la \tilde W^{2}\ra$\,,
$\la M\,\tilde W\ra$ and $\la M^{2}\,\tilde W^{2}\ra$\,.
Rewriting them as $\star$ polynomials,
we can replace $\la M^{2}\ra$ by $2(1-\l^{2})$ in $\mathcal A_{\l}$\,.
In this way, any element of $\mathcal A_{\l}$\,, namely PM generators,
can be written as $\star$ polynomials of  $\la M\,\tilde W\ra$ and $\la M^{2}\,\tilde W^{2}\ra$
with $\la \tilde W^{2}\ra$-dependent coefficients:
\ba
M^{\sst (r,2t)}(\tilde W)
&\sim&\la M^{2}\,\tilde W^{2}\ra^{\star t}\star
\la M\,\tilde W\ra^{\star (r-2t)}+\,\la \tilde W^{2}\ra\,\Big(
a\,\la M^{2}\,\tilde W^{2}\ra^{\star (t-1)}\star\la M\,\tilde W\ra^{\star (r-2t-2)}\nn
&&+\,b\,\la M^{2}\,\tilde W^{2}\ra^{\star(t-2)}\star
\la M\,\tilde W\ra^{\star (r-2t-4)}+\cdots \Big)+\cdots,
\ea
where $a$ and $b$ are some constants.
Here, one can notice that 
the $\star$ polynomial corresponding to $M^{\sst (r,2t)}$
has maximal orders $r-2t$ and $t$ in $\la M\,\tilde W\ra$ and $\la M^{2}\,\tilde W^{2}\ra$\,, respectively.

 In $\mathcal A_{\l}$\,, 
 any $\star$ powers of  $\la M^{2}\,\tilde W^{2}\ra$ can appear
 so PM generators of any even depth are present there.
However in $\mathfrak{p}_{\ell}$\,, 
one can show using \eqref{PM covariant condition} that
the $\ell$-th $\star$ power of $\la M^{2}\,\tilde W^{2}\ra$
reduces to lower order ones: 
\be
	 \la M^{2}\,\tilde W^{2}\ra^{\star \ell}\sim
	  a\,\la \tilde W^{2}\ra\,\la M^{2}\,\tilde W^{2}\ra^{\star (\ell-1)}+\cdots+
	 \la \tilde W^{2}\ra^{\ell}\,,
\ee
with some constant $a$\,.
Therefore, the generators of depth not smaller than $2\,\ell$ become 
linear combinations of generators of lower depth.
In order to remove this redundancy,
one can impose on $\tilde W^{ab}$ the orbit condition \eqref{pm orbit}, which is equivalent to
\be
	W^{ab}=w^{a\a}\,w^{b}{}_{\a}\,,\qquad
w_{(\a_1}\!\cdot w_{\a_2}\,\cdots\,w_{\a_{2\ell-1}}\!\cdot w_{\a_{2\ell})}=0\,.
    \label{orbit W para}
\ee
Here, $W^{ab}$ denotes $\tilde W^{ab}$ satisfying the above condition.
With this, any higher than $\ell-1$ $\star$-powers of $\la M^{2}\,W^{2}\ra$ vanish in $\mathfrak{p}_\ell$\,:  
\be
	 \la M^{2}\,W^{2}\ra^{\star n}=0
	 \qquad
	 [\,n=\ell,\ell+1,\ldots\,]\,.
\ee
It will be convenient to require another condition  on $W^{ab}$\,,
\be
	\la W^{2}\ra=0
	\qquad \Leftrightarrow
	\qquad
	w_\a{}\cdot w_{\g}\,w^{\g}\cdot w_\b=0\,,
	\label{W2=0}
\ee 
to hold. Note that this requirement does not project any component of 
generators in \eqref{M W} so we do not loose any generality by imposing it.
To summarize, with \eqref{orbit W para} and \eqref{W2=0},
any elements of $\widetilde{hs}(\mathfrak{so}_{d+2})$
are projected to the generators of $\mathcal{A}_\l$ (or $\mathfrak{p}_\ell$) in a
simple manner as
\be
    M^{\sst (r,2t)}(W)=\left\{\begin{array}{ccc}
    &\la M^{2}\,W^{2}\ra^{\star t}\star
    \la M\,W\ra^{\star (r-2t)}\,,\qquad &t=0,1,2\ldots\\
    & 0\,, \qquad &t=\ell,\ell+1,\ldots
    \end{array}\right.,
\ee
where the second case only holds for $\mathfrak{p}_\ell$\,.

\section{Trace and Bilinear Form}
\label{5}
Analogously to the case of massless HS algebra, 
it will be useful to define a trace in PM HS algebras, 
from which the bi-/tri-linear forms and the structure constants can be derived.
We define the trace as the coefficient of identity, similarly to the 
massless HS algebra,
\be
\tr\left[c_0+c_a T^a\right]=c_0\,,
\label{Tr Def}
\ee
where $T^a$ denotes a generator of PM HS algebra.
The expression of trace for any element in the algebra 
can be obtained in principle by only using
the definition \eqref{Tr Def} as in the case of massless HS algebra \cite{Joung:2014qya},
and  the trace formula for PM HS algebra $\mathfrak{p}_{\ell}$
is obtained in that way for a few lower $\ell$'s in Appendix \ref{A}.

In the following, we will derive the bilinear form of 
$\mathcal A_{\l}$, that should manifest the appearance of the ideal $\mathfrak{q}_{\ell}$
and the corresponding coset $\mathfrak{p}_{\ell}$
when $\l=\ell$\,.

\subsection*{Non-Gaussian projector}

In the massless case, it turns out \cite{Joung:2014qya} that the trace obtained from the definition \eqref{Tr Def}
admits a simple representation,
 \be
	\tr_{\mathfrak{p}_{1}}[f]=(\Delta\star f)(0)\,,
	\label{trace}
\ee
where $\Delta(y)$ is the \emph{projector} introduced in \cite{Vasiliev:2001wa,Vasiliev:2004cm}\footnote{Strictly speaking it is not a projector, since $\Delta\star\Delta$ is not well defined as shown in \cite{Vasiliev:2004cm}. See also recent discussion in the context of $(A)dS_2$ algebras in \cite{Alkalaev:2014qpa}.}.
The function $\D(y)$ giving the trace is not unique as shown in \cite{Joung:2014qya}
even though the definition \eqref{Tr Def} is unambiguous.

Generalizing this massless result to PM one, we assume that the 
trace of the PM algebra $\mathcal A_{\lambda}$ admits also an analogous expression,
\be
	\tr_{\mathcal A_{\lambda}}[f]
	=(\Delta_{\lambda}\star f)(0)\,,
	\label{trace lambda}
\ee
with the conditions
\be
	[\,\D_{\l}\,\overset{\star},\,K_{\a\b}\,]=0=
	[\,\D_{\l}\,\overset{\star},\, M_{ab}\,]\,.\label{commutation condition}
\ee
The above conditions are satisfied by the ansatz,
\be
\D_{\l}=\D_{\l}(z)\,,\qquad z=K_{\a\b}\,K^{\a\b}\,.
\ee
In the massless case, we also impose the condition that
$K_{\a\b}\star \D_{1}(z)=0$\,.
In the case of $\mathcal A_{\l}$\,, we impose instead
\ba\label{l-1condition}
\left(
C_2(\mathfrak{sp}_2)-(1-\lambda)(1+\lambda)\right)
\star\Delta_{\lambda}(z)=0\,.\label{c2condition}
\ea
In order to determine such $\D_{\l}$\,, we assume
\be
	\Delta_{\lambda}(z)=\sum_{n=1}^{\infty}\,d_{n}(\lambda)\,\Pi_{n}(z)\,,
	\label{projector def}
\ee
with $\Pi_{n}(z)$ satisfying
\be\label{covariant projector definition}
K_{(\a_1\a_2}\star K_{\a_3\a_4}\star\cdots\star K_{\a_{2n-1}\a_{2n})}
\star \Pi_{n}(z)=0\,.
\ee
We shall show below that \eqref{c2condition} 
fixes uniquely the coefficients $d_{n}(\l)$ and 
determines the projector.
Since
the condition \eqref{PM covariant condition}
should hold for $\l=\ell$\,, 
the coefficients in \eqref{projector def} will satisfy $d_{n\ge\ell}(\ell)=0$\,.
The identification of $\Pi_{n}(z)$ is in parallel with $\D_{1}$ in \cite{Vasiliev:2004cm},
and the equation \eqref{covariant projector definition} for $\Pi_{n}$ 
takes the following form,
\be\label{Delta equation}
\left(2\,z\,\partial_z^2+(d+1)\,\partial_z+1\right)^{n}\Pi_{n}(z)=0\,.
\ee
Clearly $\Pi_{m}(z)$ with $m<n$ also solves the equation for $\Pi_{n}$
but since we are considering a linear combination of them,
without a loss of generality 
we can take a particular solution of \eqref{Delta equation},
\be\label{Delta cov}
\Pi_{n}(z)=\int_{-1}^{1}ds\,(1-s^2)^{\frac{d}2-n}\,e^{i\,s\,\sqrt{2\,z}}\,,
\ee
for which $\left(2\,z\,\partial_z^2+(d+1)\,\partial_z+1\right)^{n-1}\Pi_{n}(z)\neq 0$.
With the above $\Pi_{n}(z)$'s, 
the condition \eqref{c2condition}
fixes the coefficients $d_{n}(\lambda)$ in the formula \eqref{projector def} as
\be
d_n(\lambda)=c\,\frac{(1-\l)_{n-1}\,(1+\l)_{n-1}}{(\frac32)_{n-1}\,(n-1)!}\,
\frac1{d-2n}\,,
\label{cl}
\ee
up to an overall factor $c$ which can in turn be fixed from the 
normalization condition $\tr(1)=1$ 
(or, equivalently $\Delta_{\l}(0)=1$).
Notice that the coefficient $d_{n}(\l)$ is ill-defined for $d=2n$\,.
This problem is in fact an artifact of non-Gaussian projector.
In terms of Gaussian projector, this factor will disappear 
and in the end all formulas are well-defined.
Details of computation for \eqref{cl} can be found in Appendix \ref{A}\,.
Remark also that 
the dependence of $\l$ and $d$ in \eqref{cl}
is such that there exist 
special values of $\l$ where 
the series 	\eqref{projector def}
 acquire special properties.
These special cases deserve more careful analysis since
they may imply the appearance of an ideal in the algebra $\mathcal A_{\l}$\,.
For concrete treatment, we shall explicitly calculate the 
bilinear form using the trace formula. 
However, as in the massless case, the 
non-Gaussian dependence on oscillators in the projector
 complicates the computations.
Fortunately,
we can find an alternative projector which gives
the same trace but admits a treatable expression.

\subsection*{Gaussian projector}

The integral form \eqref{Delta cov} of $\Pi_{n}$ 
is simple but not very useful for
actual computations as it involves
\be
	\exp\left(i\,s\sqrt{\frac{y_{\a}\cdot y_{\b}\,y^{\a}\cdot y^{\b}}2}\right).
\ee
However, using the same method as the one described in the Appendix of \citep{Joung:2014qya}, one can show that the expression
\be\label{Projector1}
P_{n}=
\int_0^1dx\, x^{\frac12}\, (1-x)^{\frac{d-2}2-n}\, e^{-2\,\sqrt{x}\,y_+\cdot\, y_-}\,
\ee
is equivalent to $\Pi_{n}$\, in the following sense:
\be
	(\Pi_{n}\star f)(0)=(d-2n)\,(P_{n}\star f)(0)\,.
\ee
Here, one can see that the problematic factor $d-2n$ cancels out
when working with the Gaussian projector.
Since $\Delta_\l$ will be used only within the setting of \eqref{trace lambda}, we can replace $\Pi_n$ by $(d-2n) P_n$ in the formula of the projector.
Evaluating the sum in \eqref{projector def}, we obtain 
new projector as
\be\label{Projector}
	D_{\l}=N_{\l} \int_0^1 dx\, x^{\frac12}\, (1-x)^{\frac{d-4}2}\,
	{}_2F_1\left(1+\l\,,\, 1-\l\,;\, \frac32\,;\, \frac1{1-x}\right)
	e^{-2\,\sqrt{x}\,y_+\cdot\,y_-}\,,
\ee
with
\be
N_{\l}=\frac{(-1)^{\l-1}\,\Gamma(d+1)}{2^{d-1}\,
\Gamma(\frac{d}2-\l)\,\Gamma(\frac{d}2+\l)}\,.
\label{NG projector}
\ee
Let us emphasize again that this projector is equivalent to $\Delta_\l$
since 
\be 
    \tr(f)=\left(\Delta_\l \star f\right)(0)=
    \left(D_\l \star f\right)(0)\,,
    \qquad \forall\,f\in \widetilde{hs}(\mathfrak{so}_{d+2})\,.
    \label{trace D}
\ee
The apparent advantage of this projector is that it involves
a Gaussian function of oscillators and therefore is convenient for actual calculations.
Hence, we will use the expression \eqref{Projector}
in the computation of invariant bilinear form of the PM HS algebra. 
The trace defined with \eqref{Projector}
can be equally obtained starting from the definition \eqref{Tr Def}
as we have demonstrated for $\mathfrak{p}_{\ell}$ 
in Appendix \ref{A}.

\subsection*{Bilinear form}

Analogously to massless case \cite{Joung:2014qya}, we consider now the invariant bilinear form of the algebra given as a trace from star-product of generating functions of all PM generators:
\be
B(W_1,W_2)=\tr(M(W_1)\star M(W_2))
=\tr\left(e^{y_+\cdot W_1\cdot y_-}\star e^{y_+\cdot W_2\cdot y_-}\right).
\label{bilin}
\ee
We use \eqref{trace D} for the trace, 
hence consider first
\be
\label{starrho}
\frac1{G^{\sst (2)}(\rho,W)}=e^{\rho\, y_{+}\cdot\,y_{-}}\star e^{y_+\cdot W_1\cdot y_-}\star e^{y_+\cdot W_2\cdot y_-}\,\big|_{y_{\a}=0}\,,
\ee
which  can be evaluated using the 
star product composition formula derived in \cite{Didenko:2003aa}
and used in \citep{Joung:2014qya} as
\ba\label{G2}
G^{\sst (2)}(\rho,W)=\frac{\det_{\sst N\times N}\!\left[\frac12\,\frac{1+\rho}{1-\rho}
\,\frac{2+W_1}{2-W_1}\,\frac{2+W_2}{2-W_2}+\frac12\right]}
{\det_{\sst N\times N}\!\left[\frac12\,\frac{1+\rho}{1-\rho}+\frac12\right]
\det_{\sst N\times N}\!\left[\frac12\,\frac{2+W_1}{2-W_1}+\frac12\right]
\det_{\sst N\times N}\!\left[\frac12\,\frac{2+W_2}{2-W_2}+\frac12\right]}\,.
\ea
This expression involves determinants of
$N\times N$ matrices, which can be simplified to
\be
G^{\sst (2)}(\rho,W)
=\det{\!}_{\sst N\times N}\!\left[1+\rho\,\frac{W_1+W_2}2+\frac{W_1\,W_2}4\right].
\ee
Further simplification can be made by
defining the \emph{dual} matrices $V_{ij}$ as
\be
	(V_{ij})_{\a\b}=(w_i)_{\a}{}^{A}\,(w_{j})_{\b A}
	\qquad
	\left[\,(W_{i})^{AB}=(w_{i})^{\a A} (w_{i})_{\a}{}^{B}\,\right],
\ee
and this enables us to rewrite (using Sylvester's determinant theorem) the $N\times N$ determinant
as $4\times 4$ one as
\be
G^{(2)}(\rho,W)=\det{}_{\sst 4\times 4}\!
\begin{bmatrix}
  1+\frac{\rho}2\,V_{11}+\frac14\,V_{12}\,V_{21}  &\quad \frac{\rho}2\,V_{12}+\frac14\,V_{12}\,V_{22} \\
    \frac{\rho}2\,V_{21}       					  & 1+\frac{\rho}2\,V_{22}
\end{bmatrix}.
\ee
Using the parameterization \eqref{orbit W para} and 
the condition \eqref{W2=0},
the above reduces to
\be
G^{\sst (2)}(\rho,W)=
\det{\!}_{\sst 2\times 2}\left[1+\frac{1-\rho^2}4\,V_{12}\,V_{21}
+\frac{\rho}2\,V_{11}-\frac{\rho\,(1-\rho^2)}8\,
V_{12}\,V_{22}\,V_{21}\right],
\ee
which can be straightforwardly evaluated to give
\be\label{G2PM}
G^{\sst (2)}(\rho,W)=
\left(1+\frac{1-\rho^2}8\,\la W_1\,W_2 \ra\right)^2
+\frac{\rho^2\,(1-\rho^2)}{16}\,\la W_1^2\,W_2^2 \ra\,.
\ee
Finally, using \eqref{Projector}, \eqref{starrho} and \eqref{G2PM},
the invariant bilinear form can be expressed as 
\ba\label{Bilinear}
B(W_1,W_2)
=N_{\l}\,\int_0^1 dx\ \frac{x^{\frac12}\,
(1-x)^{\frac{d-4}2}\, {}_2F_1(1+\l,\ 1-\l;\ \tfrac32;\ \tfrac1{1-x})}
{\left(1+\frac{1-x}8\,\la W_1\,W_2\ra\right)^2
+\frac{x\,(1-x)}{16}\,\la W_1^2\,W_2^2\ra}\,.
\ea
Let us make a few observations on the bilinear form \eqref{Bilinear}.
First of all, it is manifestly symmetric in $\lambda \to -\l$, which is a straightforward consequence of the symmetry of the quadratic Casimir.
Moreover, when $\l=\ell$ 
the bilinear form become degenerate for the generators 
$M^{\sst (r,2t)}$ with $t\ge \ell$\,.
This degeneracy is less manifest since 
$\la W_{1}\,W_{2}\ra^{r-2t}\,\la W_{1}^{2}\,W_{2}^{2}\ra^{t}$
does not exactly correspond to the contribution of $M^{\sst (r,2t)}$
--- in other words, the bilinear form is not diagonal.
Taking the simplest example of $\l=\ell=1$ case,
the $W^{2}$-order term of the bilinear form is given by
\ba
	&&
	a\left(\la W_{1}\,W_{2}\ra^{2}
	-\frac{4}{d}\,\la W_{1}^{2}\,W_{2}^{2}\ra\right)
	=W_{1}^{a_{1}b_{1}}\,W_{1}^{a_{2}b_{2}}\,
	W_{2}^{c_{1}d_{1}}\,W_{2}^{c_{2}d_{2}}\times\nn
	&&\qquad \times
	\left[\tr\left(M^{\sst (2,0)}_{a_{1}a_{2},b_{1}b_{2}}\,
	M^{\sst (2,0)}_{c_{1}c_{2},d_{1}d_{2}}\right)
	+\tr\left(M^{\sst (2,2)}_{a_{1}a_{2}}\,M^{\sst (2,2)}_{c_{1}c_{2}}\right)
	\eta_{b_{1}b_{2}}\,\eta_{d_{1}d_{2}}\right],
\ea
with some constant $a$\,.
By decomposing the LHS of the equation into traceless
tensors, one can verify that the PM generators of the second depth
have vanishing bilinear from:
\be
	\tr\left(M^{\sst (2,2)}_{a_{1}a_{2}}\,M^{\sst (2,2)}_{c_{1}c_{2}}\right)=0\,,
\ee
hence form  an ideal.
Likewise, the appearance of the ideal $\mathfrak{q}_{\ell}$ and the
corresponding coset algebra $\mathfrak{p}_{\ell}$ in $\mathcal A_{\ell}$
can be checked with diagonalization. And as we discussed before,
 the ideal part --- having vanishing bilinear form --- can be
 conveniently discarded by contracting with $W^{ab}$ satisfying
 the orbit condition \eqref{pm orbit}.
 We shall keep this condition in the following section.

\section{Finite dimensional truncations}
\label{6}
The normalization factor $N_{\l}$ \eqref{NG projector} of the bilinear form \eqref{Bilinear}
contains 
\be
	\frac1{\Gamma(\frac d2-\l)}\,,
\ee
which vanishes for $\l-\frac d2=0,1,\ldots$\,.
Remind that $N_{\l}$ is fixed with the condition
$\tr(1)=1$ which means the norm of the identity 
is one. When $N_{\l}$ itself
vanishes, the only way to keep the normalization of the identity
is that the integral \eqref{Bilinear} should diverge 
for the identity part.
In this case, if there exists generators which give 
finite integral, then their norm will vanish
and  the algebra develops a new ideal corresponding
to such generators.

In order to study these new ideals and corresponding coset algebras, 
it is useful to use the transformation of hypergeometric function,
\be
{}_2F_1(1+\l,\,1-\l;\,\tfrac32;\,\tfrac1{1-x})=
(\tfrac{x}{x-1})^{\l-1}{}_2F_1(\tfrac12-\l,\,1-\l;\,\tfrac32;\,\tfrac1{x})\,,
\ee
to rewrite the bilinear form as
\ba\label{Bilinear odd d}
B(W_1,W_2)
=N_{\l}\,(-1)^{\l-1}\int_0^1 dx\, 
\frac{x^{\l-\frac12}\ (1-x)^{\frac{d-2}2-\l}\, 
{}_2F_1\left(\frac12-\l\,,\, 
1-\l\,;\, \frac32\,;\, \frac1{x}\right)}
{\left(1+\frac{1-x}8\,\la W_1\,W_2\ra\right)^2
+\frac{x\,(1-x)}{16}\,\la W_1^2\,W_2^2\ra}\,.
\ea
Focusing on the $M^{\sst (r,2t)}$ part, which corresponds to
\be
	\la W_1\,W_2\ra^{r-2t}\,\la W_1^2\,W_2^2\ra^{t}\,,
	\label{Wr Wt}
\ee
the integral is proportional, up to finite coefficient, to
\ba
&&\int_0^1 dx\, x^{\l-\frac12+t}\,
(1-x)^{\frac{d-2}2-\l+r-t}\,
{}_2F_1\left(\tfrac12-\l\,,\, 1-\l\,;\, \tfrac32\,;\, \tfrac1{x}\right)\nn
&&=\,\Gamma\left(\frac{d}2-\l+r-t\right)
\sum_{n=0}^{\infty}\,\frac{(\frac12-\l)_{n}\,
(1-\l)_{n}\,\Gamma(\l+t+\frac12-n)}
{n!\,(\frac32)_{n}\,\Gamma(\frac{d+1}2+r-n)}\,.
\label{integral}
\ea
This integral diverges when the first factor $\Gamma(\frac d2-\l+r-t)$ diverges 
while the summation part is always finite.
Hence, for a fixed value of $\l$ with vanishing $N_{\l}$\,:
\be
\l=\frac d2+k\,,\qquad k=0,1,\ldots
\label{trunc lambda}
\ee
the algebra develops an infinite-dimensional ideal corresponding to
the generators $M^{\sst (r,2t)}$ with $r-t>k$\,. The coset algebra,
denoted henceforth by $\mathfrak{f}_{k}$\,,
 is finite dimensional one with the generators $M^{\sst (r,2t)}$ satisfying
\be
	r-t\le k\,.
\ee
The Young diagram of $M^{\sst (r,2t)}$ has $r$ 
and $r-2t$ boxes in the first and second row respectively,
so  the total number  is $2(r-t)$\,.
Therefore, the algebra $\mathfrak{f}_{k}$ consists
of the generators whose Young diagram contains no more than $2k$ boxes.
%
This situation is very similar to the one with massless HS algebras in three dimensions\footnote{In fact, the three dimensional HS algebra $hs[\l]$ can be interpreted as the partially-massless HS algebra in $(A)dS_2$, and allows for finite-dimensional truncations, interpreted as partially-massless algebras \cite{Alkalaev:2014qpa}.}, and five dimensions \cite{Joung:2014qya,Manvelyan:2013oua}  where the corresponding HS algebras  are the symmetries of symmetric tensor representations of $\mathfrak{sl}_4$ or, equivalently, (anti-)self-dual (spin-)tensors of $\mathfrak{so}_6$ described by rectangular three-row Young diagrams.
For example, $\mathfrak{f}_{k}$ with $k=3$ have (we omit the identity generator, corresponding to Young diagram with no boxes)
\ba
&\left\{\ \small\yng(1,1)\ ,\ \small\yng(2)\ \right\},\qquad
\left\{\ \small\yng(2,2)\ ,\ \small\yng(3,1)\ , \ \small\yng(4)\ \right\},&\nn
&\left\{\ \small\yng(3,3)\ ,\ \small\yng(4,2)\ ,
\ \small\yng(5,1)\ ,\ \small\yng(6)\ \right\},&
\ea
where we have organized the generators in terms of the number of boxes.
 Regrouping the same set according to spins, we get 
\ba
\left\{\ \small\yng(1,1)\ \right\},\qquad
\left\{\ \small\yng(2,2)\ ,\ \small\yng(2)\ \right\},\qquad
\left\{\ \small\yng(3,3)\ ,\ \small\yng(3,1)\ \right\},\qquad \nn
\left\{\ \small\yng(4,2)\ ,\ \small\yng(4)\ \right\},\qquad
\left\{\ \small\yng(5,1)\ \right\},\qquad  
\Big\{\ \small\yng(6)\ \Big\}\,.
\ea
The dimension of $\mathfrak{f}_{3}$ is 
$\left(\tfrac{(d+1)(d+2)(d+6)}6\right)^{2}$\,.
For generic integer value of $k$\,, 
the dimension of $\mathfrak{f}_{k}$ is given again by  a perfect square $M_{k}^{2}$ with
\be
	M_{k}=\frac{(d+1)_{k-1}(d+2k)}{k!}\,.
\ee
This suggests that the algebra $\mathfrak{f}_{k}$ is isomorphic 
to $\mathfrak{gl}_{M_{k}}$\,,
an endomorphism of a $\mathfrak{so}_{d+2}$ representation
with dimension $M_{k}$\,. 
It turns out that this representation corresponds to
the finite-dimensional one-row $\mathfrak{so}_{d+2}$ Young diagram 
with length $k$\,:
\be
	\begin{tikzpicture}
	\draw (0,0) rectangle (4,0.4);
	\draw [very thin, gray] (0.4,0) -- (0.4,0.4);
	\draw [very thin, gray] (3.6,0) -- (3.6,0.4);
	\node at (2,0.2) {$k$};
	\end{tikzpicture}\,.
	\label{fin k rep}
\ee
This can be verified by comparing the quadratic Casimir of \eqref{fin k rep}
--- given by $k(k+d)$ ---
with that of \eqref{trunc lambda}\,: the latter gives
\be
	C_{2}(so(d+2))=-\frac{(d-2\,\lambda)(d+2\,\lambda)}{4}
	=k(k+d)\,,
\ee
hence they coincide as expected.
This proves that the finite-dimensional coset algebra $\mathfrak{f}_{k}$
corresponds to the symmetries of the finite-dimensional representation
\eqref{fin k rep}. 
The latter representation can be also understood from 
the boundary point of view: 
they should correspond to the solution (sub)space of higher-order Laplace equation for scalar field $\Box^{d/2+k}\phi=0$\,.
Since any higher-order Laplace equation, $\Box^{\ell}\phi=0$\,, 
is equivalent to the set of equations in the $(d+2)$-dimensional ambient space:
\be 
    \partial_X^2\,\Phi(X)=0\,,\qquad
    \left(X\cdot \partial_X+\frac{d}2-\ell\right)\,\Phi(X)=0\,,
    \label{amb sys}
\ee 
one can study its solutions using the above system (see e.g. \citep{Bekaert:2013zya})\footnote{In \eqref{amb sys},
we can consider the transformation
\be
	\Phi(X)\quad\to\quad (X^{2})^{\ell-d}\,\Phi(X)\,,
\ee
to flip the sign of $\ell$\,. Higher order singletons
are usually described with the positive sign --- that is, with homogeneity 
$-(d+2\ell)/2$ --- together with $(X^{2})^{\ell}\,\Phi(X)=0$\,. Equivalent description is given by equation $(\partial_X^2)^{\ell}\Phi(X)=0$ supplemented with condition $(X^2)\,\Phi(X)=0\,$. These conditions can be solved to get higher order conformal wave operators (GJMS operators) in arbitrary background \cite{Manvelyan:2007tk}.}.
For $\ell=\frac d2+k$\,, the ambient field $\Phi$ has homogeneity 
degree $k$\,, any $k$-th order polynomials in $X^a$ dual to 
traceless tensors \eqref{fin k rep} become solutions of the system \eqref{amb sys}.

Let us conclude this section with a short summary. 
In this section,  using the bilinear form \eqref{Bilinear odd d}, we have identified the finite dimensional PM HS algebras  corresponding to the symmetry of the symmetric tensor representation \eqref{fin k rep}  of (A)dS algebra.  
Retrospectively,  these algebras could be identified with no reference to the bilinear form but
only scanning possible finite dimensional representations of (A)dS algebra whose tensor square
satisfy the condition of PM algebras.

\section{PM Algebras and CHS Symmetries}
\label{7}

Conformal Higher Spin (CHS) theory is an interacting theory of conformal higher-spin fields
 ---
that is, defined by Fradkin-Tseytlin free action \cite{Fradkin:1985am} --- 
of all spins \cite{Segal:2002gd,Bekaert:2010ky} and its symmetry coincides
with the symmetry of massless
HS theory in one higher dimensions.
It has been shown that $(d+1)$-dimensional free conformal spin-$s$  field 
can be decomposed, around $(A)dS_{d+1}$\,, into the set of spin-$s$ PM fields 
with all depths \cite{Joung:2012qy,Metsaev:2014iwa,Nutma:2014pua}: 
\be
	{\rm CHS}_{s}=\bigoplus_{t=0}^{s+\frac{d-5}2} {\rm PM}_{(s,t)}\,,
\ee
where ${\rm PM}_{(s,t)}$ with $t\ge s$ are massive fields.
Therefore, CHS symmetry itself can be regarded as the symmetry of PM fields of any depths.
Then a natural question arises: whether the even depth PM symmetries 
discussed in this paper ($\mathcal{A}_{\l}, \mathfrak{p}_{\ell}$ or $\mathfrak{f}_{k}$)
can be embedded as a subalgebra in CHS symmetry. 
First of all, it is rather straightforward to see that the even depth PM part of CHS symmetry form a subalgebra. 
CHS symmetry in $(A)dS_{d+1}$ is isomorphic to  the massless HS algebra 
in $(A)dS_{d+2}$\,, so can be realized by $d+2$ sets of oscillators, $Y_{\a A}=(y_{\a a},z_{\a})$\,. Now considering the map, 
\be  
	\rho\,:\ \left(\,y_{\a a}\ ,\, z_\a\,\right)
	\quad \to \quad \left(\,y_{\a a}\ ,\,  -z_\a\,\right),
\ee
which is an automorphism
 for $hs(\mathfrak{so}_{d+3})$
the  $\rho$ invariant space 
of $hs(\mathfrak{so}_{d+3})$ forms a subalgebra
and such space is generated by even depth PM generators.

The next question is to what this symmetry corresponds. 
One of the simplest way to answer this question is 
by examining the Howe duality in the oscillator construction.
The quotient relation of massless HS algebra in $AdS_{d+2}$
can be translated in $AdS_{d+1}$ to
\be
	K_{\a\b}+z_{\a}\,z_{\b}\sim 0\,,
\ee
where $K_{\a\b}$ is the $\mathfrak{sp}_{2}$ generators dual
to $\mathfrak{so}_{d+2}$ --- not $\mathfrak{so}_{d+3}$\,.
The above relation simply means that the dual $\mathfrak{sp}_{2}$
carries the representation realized by $z_{\a}\,z_{\b}$\,.
By calculating the quadratic Casimir, we get
\be
	C_{2}(\mathfrak{sp}_{2})\sim \frac12\,z_{\a}\,z_{\b}\star z^{\a}\,z^{\b}
	=\frac{3}{4}\,,
\ee
and can recognize that it coincides with the $C_2$ of  $\mathcal A_{\frac12}$\,.
This proves that $\mathcal A_{\frac12}$ is the even depth subalgebra of CHS symmetry.

The existence of even depth PM subalgebra $\mathcal A_{\frac12}$ inside of CHS symmetry has an interesting implication towards a unitary truncation of CHS theory.
CHS theory is considered to be non-unitary since its linearized spectrum is described by 
a higher derivative action.
The non-unitarity of the latter become clear when it is decomposed into the actions of PM fields around $(A)dS_{d+1}$:
\be
	S^{\sst\rm CHS}_{s}=\sum_{t=0}^{s+\frac{d-5}2} (-1)^{t}\,
	S^{\rm\sst PM}_{(s,t)}\,,
	\label{CHS PM}
\ee
where even depth PM fields and odd depth PM fields have relatively negative sign for kinetic terms.
One may wonder whether the above CHS action
can be truncated into a unitary one by simply selecting 
positive sign part.
Although it is not certain whether this truncation might be consistent, 
one can immediately identify one necessary condition for 
the consistency: the symmetry of the truncated
spectrum should form a subalgebra of the original symmetry.
For example,
the spectrum of Conformal Gravity (CG) 
decomposes into massless spin two and PM spin two,
and their kinetic terms have relatively negative sign.
The symmetry of CG --- that is the conformal symmetry ---
contains the symmetry of Einstein Gravity --- that is the isometry ---
as a subalgebra,
and the action of CG 
written as \eqref{CHS PM}
can be consistently truncated to Einstein Hilbert one
\cite{Metsaev:2007fq,Maldacena:2011mk,Deser:2012qg}. 
Now coming back to CHS theory, if CHS action can be 
truncated into a theory of even depth PM fields,
we can escape from the non-unitarity problem 
at least at the linear level. 
The fact that $\mathcal A_{\frac12}$ 
corresponds to a subalgebra of CHS symmetry suggests
that such a truncation 
might be viable. 

Another interesting question is that whether other even depth PM symmetries we have encountered in this paper can
be still considered as a truncation of (a variant of) CHS theory. 
Although we do not have a proof, we do not see how the other PM algebras can be embedded in CHS symmetry.
Moreover, we do not see even the massless HS symmetry,
that is $\mathfrak{p}_{1}$\,, can be so:
interpreted differently, we do not see
how usual HS symmetry can be
embedded inside of the same symmetry in one higher dimensions.
Considering the embedding of $\mathfrak{p}_{\ell}$ within conformal-like symmetry,
there is a chance that $\mathfrak{p}_{\ell}$ is supplemented with odd depth generators 
to form a new symmetry.
Corresponding theories may coinside with Weyl-like theories of partially-massless fields of both even and odd depth lower than a given number that were considered in \cite{Joung:2012qy}. Conversely, Weyl-like theories of \cite{Joung:2012qy} may have even-depth unitary truncations that make use of the algebras $\mathfrak{p}_{\ell}$.

\acknowledgments

We thank S. Rey  for useful discussions. 
The work of EJ was supported by Basic Science Research
Program through the National Research Foundation of Korea (NRF) funded by the Ministry of Education (2014R1A6A3A04056670)
and by the Russian Science Foundation grant 14-42-00047 associated with Lebedev Institute. 
The work of KM was supported by the BK21 Plus Program funded by the Ministry of Education (MOE, Korea) and National Research Foundation of Korea (NRF).
\appendix

\section{Computational Details}\label{A}

\subsection*{Gaussian Projector}

We derive here the trace of PM HS algebra without initial assumption on the existence of a trace projector. We follow the same route 
as the derivation of the trace formula for massless HS algebras in \cite{Joung:2014qya}.
First, we compute
\be
	\tr\left[ \exp\left(y_{-}\!\cdot \tilde W\cdot y_{+}\right)\right]=
	t_{\ell}\left(\la \tilde W^{2}\ra\right).
\ee
By taking the maximal trace of the above, we get 
\be
	t_{\ell}(z)=\sum_{n=0}^{\infty} \t_{\ell,n}\,
	\frac{z^{n}}{\left(\frac {d+2}2\right)_{n}\,\left(\frac{d+1}2\right)_{n}}\,,
	\qquad
	\t_{\ell,n}=
	\bigg[\hspace{-3.5pt}\bigg[
	\left(\frac{y_{+}\!\cdot y_{[-}\,y_{+]}\!\cdot y_{-}}4\right)^{n}
	\bigg]\hspace{-3.5pt}\bigg]\,.
\ee
In order to compute the coefficients $\t_{n}$\,,
we use \eqref{PM equiv}
and  obtain
the following recurrence relations,
\ba
	&&
	\t_{\ell,n+1}-\frac18\left(n+\frac32\right)
	\left(n+\frac {d+2}2\right) \t_{\ell,n}
	+\frac{\ell+1}4\s_{\ell,n}=0\,,\nn
	&&
	\s_{\ell,n+1}-\frac18\,(n+1)
	\left(n+\frac {d+1}2\right) \s_{\ell,n}
	+\frac{\ell-1}2\,\t_{\ell,n+1}=0\,,
\ea
where
\be
\s_{\ell,n}:=\bigg[\hspace{-3.5pt}\bigg[ \frac{y_{+}\!\cdot y_{-}}2
	\left(\frac{y_{+}\!\cdot y_{[-}\,y_{+]}\!\cdot y_{-}}4\right)^{n}
	\bigg]\hspace{-3.5pt}\bigg]\,.
\ee
Defining $\t_{\ell,n}$ as
\be
	\t_{\ell,n}=p_{\ell}(n)\,\t_{1,n}\,,\qquad
	\t_{1,n}=\frac{\left(\frac32\right)_{n} \left(\frac {d+2}2\right)_{n}}{8^{n}}\,,
\ee
the solutions for $p_{\ell}(n)$ read for a few low $\ell$'s
\ba
	&&p_{2}(x)=\frac{d+2+4\,x}{{d+2}}\,,
	\qquad p_{3}(x)=\frac{3(d+2)(d+4)+32\,(d+2)\,x+64\,x^{2}}{3\,(d+2)(d+4)}\,,\nn
	&&p_{4}(x)=\frac{(d+2+4\,x)[(d+4)(d+6)+16(d+2) x+32\,x^{2}]}
	{(d+2)(d+4)(d+6)}\,.
\ea
Using all these results, one can show that 
\be
	t_{\ell}(z)=p_{\ell}(z\,\partial_{z})\,t_{1}(z)\,,
	\label{t ell}
\ee
where $t_{1}(z)$ has been determined in \cite{Joung:2014qya} as
\be
	t_{1}(z)
	={}_{2}F_{1}\big(\,1\,,\,\tfrac32\,;\,\tfrac{d+1}2\,;\,\tfrac z8\,\big)
	=
	\frac{\Gamma\!\left(\frac {d+1}2\right)}{\Gamma\!\left(\frac32\right)
	\Gamma\!\left(\frac{d-2}2\right)}\,
	\int_{0}^{1} dx\,\frac{x^{\frac12}\,(1-x)^{\frac{d-4}2}}{1-x\,\tfrac z8}\,.
\ee
From \eqref{t ell} and performing the integration by part, we get 
\be
	t_{\ell}(z)=  \frac{\Gamma\!\left(\frac {d+1}2\right)}{\Gamma\!\left(\frac32\right)
	\left(\frac{d+2}2\right)_{\ell-1}
	\Gamma\!\left(\frac{d}2-\ell\right)}\,
	\int_{0}^{1} dx\,x^{\frac12}\,(1-x)^{\frac{d}2-\ell}\,\frac{\bar p_{\ell}(x)}{1-x\,\tfrac z8}\,,
\ee
where $\bar p_{\ell}$ are given for a few lower $\ell$'s by
\be
	\bar p_{2}(x)=1+x\,,\quad
	\bar p_{3}(x)=\left(1+\tfrac13\,x\right)\left(1+3\,x\right),\quad
	\bar p_{4}(x)=(1+x)\left(1+6\,x+x^{2}\right).
\ee
Finally, the formula for the trace can be put as
\be
	\tr\big[\,f(y)\,\big]
	=(f \star D_{\ell} )(0)\,,
	\label{trace1}
\ee
with $D_{\ell}$ given by 
\be
	D_{\ell}(y)= 
	\frac{\Gamma\!\left(\frac {d+1}2\right)}{\Gamma\!\left(\frac32\right)
	\left(\frac{d+2}2\right)_{\ell-1}
	\Gamma\!\left(\frac{d}2-\ell\right)}\,
	\int_{0}^{1} dx\,x^{\frac12}\,(1-x)^{\frac{d-2}2-\ell}\,\bar p_{\ell}(x)\,
	e^{-2\sqrt{x}\,y_{+}\cdot\,y_{-}}\,.
	\label{D fn}
\ee
One can check that \eqref{D fn} 
coincides with \eqref{Projector} for lower $\ell$ examples provided.

\subsection*{Non-Gaussian Projector}

We sketch the derivation of the solution to \eqref{Delta equation}.
First we perform a change of variables $z=u^2/2$
to rewrite  the equation as
\ba
\big[\tfrac{1}{u}\,\mathcal{L}_u\big]^{n}\,\Pi_{n}(u)=0,\quad\
\mathcal{L}_u=u\,\partial_u^2+d\,\partial_u+u\,,
\ea
which
is equivalent to the recursive differential equation:
\be
\mathcal{L}_u\,\Pi_{n}(u)= u\,\Pi_{n-1}(u)\,.
\ee
We move to the Fourier space
where the differential equation becomes first-order one as
\be
\Pi_{n}(u)=\int_{-1}^1 ds\, \tilde{\Pi}_{\ell}(s)\, e^{i\,s\,u}\,,
\qquad
\big[(1-s^2)\partial_s+(d-2)s\big]\,\tilde{\Pi}_{n}(s)
=\partial_s\,\tilde{\Pi}_{n-1}(s)\,,
\ee
and its solution can be identified 
with  arbitrary integration constants $a_{k}$ as
\ba
\tilde{\Pi}_{n}(s)=\sum_{k=1}^{n}\,a_k\,(1-s^2)^{\frac{d}2-k}\,.
\ea
Since we will consider the linear combination \eqref{projector def},
we can take $a_{k}=\delta_{k,n}$ without a loss of generality.

The next point we will detail here is the determination
of relative constant $d_{n}(\l)$ in \eqref{projector def}
from the condition \eqref{c2condition}.
For that, we need to compute
\be
	K^{\a\b}\star K_{\a\b}\star \Pi_{n}
	=\frac12\left[u\,\mathcal{L}_u\,\frac{1}{u}\,\mathcal{L}_u
	+6\left(\partial_u+\frac{d}{2\,u}\right)
	\mathcal{L}_u\right]\Pi_{n}\,.
\ee
With the identities, 
\be
	u\,\Pi_{n}=-(d-2n)\,\partial_{u}\,\Pi_{n+1}\,,
	\qquad
 	\partial_{u}^{2}\,\Pi_{n}=\Pi_{n-1}-\Pi_{n}\,,
\ee
one can express 
the action of the quadratic Casimir on $\Pi_{n}$ as
\ba
C_2(\mathfrak{sp}_{2})\star \Pi_{n}
&=&(1-n)(1+n)\left(
{\Pi}_{n}-\frac{(2n-1)(d-2n)}{2(n+1)(d-2n+2)}\,{\Pi}_{n-1}
\right).
\ea
With this, the condition \eqref{c2condition}
defines a recurrence relation for $d_n(\l)$ 
which can be straightforwardly solved to give \eqref{cl}.

\bibliographystyle{JHEP}

\end{document}